\documentclass[12pt]{article}
\usepackage{graphicx,wrapft}
\begin{document}
\def\lsim{\stackrel{\scriptstyle <}{\phantom{}_{\sim}}}
\def\gsim{\stackrel{\scriptstyle >}{\phantom{}_{\sim}}}

\begin{center}
{\Large\bf Nuclear pasta structures in neutron stars and the charge screening}

\vspace{1cm}


{
Toshitaka Tatsumi$^1$,
Toshiki Maruyama$^2$,
Dmitri~N.~Voskresensky$^{3,4}$,\\
Tomonori Tanigawa$^{5,2}$,
and
Satoshi Chiba$^2$
}

\vspace{0.5cm}

{
$^1$Department pf Physics, Kyoto University, Kyoto, 606-8502, Japan\\
$^2$Advanced Science Research Center, Japan Atomic Energy Research Institute, Tokai, Ibaraki 319-1195, Japan\\
$^3$Moscow Institute for Physics and Engineering, Kashirskoe sh.~31, Moscow 115409, Russia\\
$^4$Gesellschaft f\"ur Schwerionenforschung mbH, Planckstr. 1, 64291 Darmstadt, Germany\\
$^5$Japan Society for the Promotion of Science, Tokyo 102-8471, Japan
}

\end{center}

\vspace{1cm}

\begin{abstract}
Non-uniform structures of the nucleon matter 
are expected at subnuclear densities and above the nuclear density: 
they are called nuclear pastas and kaon pastas, respectively.
We numerically study these phases by means of the density functional 
theory with
relativistic mean-fields and the electric field; the electric
 field is properly taken into account.
Our results demonstrate a particular role of the charge screening
 effects on these non-uniform structures.
\end{abstract}

\section{Introduction}

\begin{wrapfigure}[20]{r}{0.63\textwidth}
\includegraphics[height=.35\textheight]{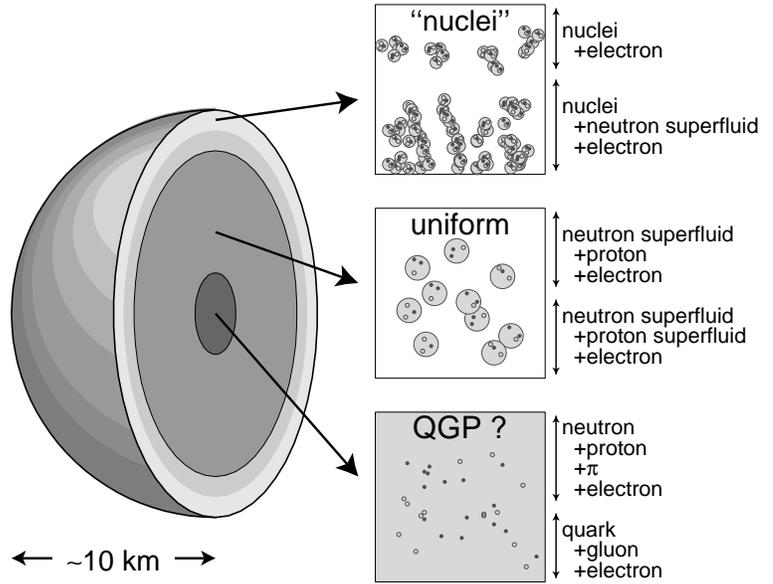}
\caption{
Schematic picture of a possible neutron star structure.
}
\label{figNstar}
\end{wrapfigure}

There are expected various form of matter inside neutron stars  
(Fig.~\ref{figNstar}),
many of which are associated with first order phase transitions (FOPT).
Recently there are many studies of the mixed phases at these FOPT such as 
hadron-quark deconfinement transition \cite{gle92,HPS93,voskre,emaru1}, 
kaon condensation \cite{GS99,CG00,CGS00,PREPL00,MYTT,RB,NR00,maruKaon}, 
color superconductivity \cite{ARRW,bed,RR04}, superfluidity in atomic traps
\cite{BCR03}, nuclear pastas
\cite{Rav83,Has84,Wil85,Oya93,Lor93,Cheng97,Mar98,Gen00,Gen02,Gen03}, etc.
Before the remark by Glendenning many authors used the Maxwell
construction (MC) to get the equation of state (EOS) 
in thermodynamic equilibrium for FOPT. Nowadays 
there exists a view that not all Gibbs conditions can be satisfied in
the description of the Maxwell construction in {\it multi-component} 
systems and the appearance of the mixed phases is inevitable, cf. 
\cite{gle92,CG00}.

In this lecture we discuss two types of the mixed phases expected in
nuclear matter; one is so called the 
``nuclear pastas'' at subnuclear densities, and the 
other is the one following kaon condensation above the nuclear density. 
We shall see non-uniform geometrical structures in both cases.
Consider a mixed-phase consisting of two phases in equilibrium, say I and II.
The system should be totally neutral but composed of many particle species 
in chemical equilibrium through weak processes. When we impose the baryon number and charge
conservation on this system, we can easily see that only two chemical potentials are independent;
one is  baryon-number chemical potential $\mu_B$ and the other is charge chemical 
potential $\mu_Q$. 
Once $\mu_B$ and $\mu_Q$ are given, all the chemical potentials of particle
species are determined.
Then  each particle density may spatially change in the 
mixed phase by the strong and electroweak interactions, but chemical 
potentials must be constant over the whole space.
The Gibbs conditions (GC) for thermodynamic equilibrium between two phases
I and II are given by 
\begin{eqnarray}
T^{\rm I}&=&T^{\rm II},~~~P^{\rm I}=P^{\rm II}\nonumber\\
\mu_B^{\rm I}&=&\mu_B^{\rm II},~~~\mu_Q^{\rm I}=\mu_Q^{\rm II}
\end{eqnarray}
in this case, where $T^i$ and $P^i$ denote temperature and pressure of
each phase, respectively.

It has been claimed that usual MC satisfies only the first three conditions, and
the final one is violated because it assumes the local charge
neutrality in each phase instead of the global charge neutrality.
A naive application  of the Gibbs conditions to separate bulk phases 
without 
the surface and the Coulomb interaction, demonstrates a broad density-region of
the structured mixed phase \cite{gle92,CG00}.
When one takes into account the geometrical structures like droplet, rod
and slab by extending the bulk calculation, one can see that the surface tension and
Coulomb interaction determine their size \cite{Rav83,HPS93,voskre}.
However, the charge screening effect (caused by the rearrangement of the
charged-particle distributions)
should be very important when the typical size is of the order of
the minimal Debye screening length in the problem.
It may  largely  affect the stability condition of the geometrical structures in the
mixed phases. 
We have been recently exploring the effect of the charge screening in the
context of the various structured mixed phases \cite{voskre,emaru1,maruKaon,maru1}.


Our aim here is to investigate the non-uniform structures in nuclear
matter numerically by means of the density functional theory with 
a relativistic mean field (RMF) model. Our framework allows one to
determine the density profiles exactly without 
any sharp boundary and the surface tension used in the bulk
calculations. It 
includes the Coulomb interaction in a proper
way and we can fully take into account the charge screening effects. 
We shall figure out how the charge screening effects induce the
rearrangement of the charged-particle distributions and thereby modify
the results obtained by the bulk calculations.

\section{Nuclear pastas}

First we consider the nuclear pastas. 
At subnuclear densities, where pressure of the uniform nuclear matter is
negative, so called ``nuclear pastas'' may appear \cite{Rav83}.
In view of the phase transition, these can be regarded as the mixed
phases following the liquid-gas phase transition in nuclear matter.
\footnote{Note that there is no uniform gas phase in nuclear matter at zero temperature.}
Stable nuclear shape may change from sphere to
rod, slab, tube and to bubble with increase of the matter density. 
Pastas are eventually dissolved into the uniform matter (liquid phase) at
a certain nucleon density below
the saturation density, $\rho_0\simeq 0.16~$fm$^{-3}$.

Existence of such pasta phases instead of the 
crystalline lattice of nuclei or the liquid $n p e$ phase would modify some important processes
by changing the hydrodynamic properties
and the neutrino opacity in the supernova matter and in the protoneutron
stars.
Also the pasta phases may influence
neutron star quakes and pulsar glitches
via the change of mechanical properties of the crust matter.

A number of authors have discussed the nuclear pastas using various models
\cite{Rav83,Has84,Wil85,Oya93,Lor93,Cheng97,Mar98,Gen00,Gen02}, but most
of them have relied on the bulk calculations;
roughly speaking, the favorable nuclear shape is determined by
a balance between the surface and the Coulomb energies.
Thus the Coulomb interaction as well as the surface tension has an crucial role in the
non-uniform pasta structures.
However, the treatment of the Coulomb interaction so far has been
rather simple and 
the rearrangement effect on the density profiles of 
the charged particles by the Coulomb interaction has been discarded in
the bulk calculations. 
The paper \cite{Gen03} discusses the effect of electron screening to 
demonstrate that it is of a minor importance, but  
the rearrangement of the proton density as a consequence of 
the Coulomb repulsion was not shown up in their study.

\subsection{Density functional theory with relativistic mean field}\label{Func}


Following the idea of the density functional theory (DFT) with
the RMF model \cite{refDFT}, we can formulate 
equations of motion to  study non-uniform nuclear matter numerically.
The RMF model with fields of mesons and baryons introduced
in a Lorentz-invariant way is simple for numerical calculations, 
but realistic enough to reproduce the bulk properties of finite nuclei
as well as saturation properties in nuclear matter.
In our framework, the Coulomb interaction is properly included in 
equations of motion for nucleons, electrons, and meson mean-fields,
and we solve the Poisson equation for the Coulomb potential $V_{\rm Coul}$
self-consistently with them.
Thus the baryon and electron density profiles, as well as the meson
mean-fields, are determined in a way fully
consistent with the Coulomb potential.


To begin with, we present the thermodynamic potential for
the neutron, proton and electron system with 
 chemical potentials $\mu_n$, $\mu_p$ and $\mu_e$, respectively; 
\begin{equation}\label{Omega-tot}
\Omega = \Omega_N+\Omega_M
    +\Omega_e,
\end{equation}
where
\begin{equation}\label{eq:OmegaN}
\Omega_N  = 
  \sum_{a=p,n}
  \int  d^3r 
  \left[
  \int_0^{k_{{\rm F},a}}
  { d^3k \over 4\pi^3}
  \sqrt{{m_N^*}^2+k^2}-\rho_a\nu_a
  \right]
  ,
\end{equation}
with the local Fermi momenta, $k_{{\rm F},a}({\bf r})(a=n,p)$, for nucleons,
\begin{eqnarray}
&&\Omega_M = \int   d^3r \left[ 
  {(\nabla\sigma)^2 + m_\sigma^2\sigma^2 \over2} + U(\sigma)
  -{(\nabla\omega_0)^2 + m_\omega^2\omega_0^2 \over2}
  -{(\nabla R_0)^2 + m_\rho^2R_0^2\over2}  \right]  ,
 \label{eq:OmegaM}
\end{eqnarray}
for the scalar ($\sigma$) and vector mean-fields ($\omega_0, R_0$) and 
\begin{equation}
\Omega_e = \int d^3r \left[
-{1\over8\pi e^2}(\nabla {V_{\rm Coul}})^2-{(\mu_e-{V_{\rm Coul}})^4\over12\pi^2}
\right],
\end{equation}
for electrons and the Coulomb potential, $V_{\rm Coul}({\bf r})$, where
$\nu_p({\bf r})=\mu_p+{V_{\rm Coul}({\bf r})}-g_{\omega N}\omega_0({\bf
r})-g_{\rho N}R_0({\bf r}),\ \
\nu_n({\bf r})=\mu_n-g_{\omega N}\omega_0({\bf r})+g_{\rho N}R_0({\bf r}),\ \
m_N^*({\bf r})=m_N-g_{\sigma N}\sigma({\bf r})$,
and the nonlinear potential for the scalar field, 
$U(\sigma)={1\over3}bm_N(g_{\sigma N}\sigma)^3+{1\over4}c(g_{\sigma N}\sigma)^4$.
Temperature $T$ is kept to be zero in the present study.

Here we used the local-density approximation (LDA) for nucleons and electrons.
Strictly speaking, LDA is meaningful only  
if the typical length of the nucleon density variation 
is larger than the inter-nucleon distance. 
To go beyond LDA we must take
into account some derivative terms with respect to particle densities,
which can be easily incorporated in the quasi-classical
manner by  the derivative expansion within
the density functional theory \cite{refDFT}.
In the case when we suppress derivative terms of nucleon densities
they follow changes of the meson mean fields and the Coulomb field that
have derivative terms. 
We must also bear in mind that for small
structure sizes, quantum effects become prominent which we disregarded.
 Here we consider large-size pasta structures and
simply discard the derivative terms, as a first-step calculation.

Parameters of the RMF model are chosen to reproduce saturation properties
of nuclear matter:
the minimum energy per baryon $-16.3$ MeV at $\rho =\rho_0 =0.153$ fm$^{-3}$,
the incompressibility $K(\rho_0) =240$ MeV, the effective nucleon  mass
$m_N^{*}(\rho_0)=0.78m_N$; $m_N =938$ MeV, and the isospin-asymmetry coefficient $a_{\rm sym}=32.5$~ MeV.
Coupling constants and meson masses used in our calculation are listed in Table 1.
\begin{table}
\caption{
Parameter set used in RMF in our calculation.
}
\begin{center}
\begin{tabular}{cccccccc}
\hline
$g_{\sigma N}$ & 
$g_{\omega N}$ &
$g_{\rho N}$ &
$b$ &
$c$ &
$m_\sigma$ [MeV]&
$m_\omega$ [MeV]&
$m_\rho$ [MeV]\\ [1mm]
\hline\\
6.3935 & 
8.7207 & 
4.2696 & 
0.008659 &
0.002421 &
 400 &
 783 &
 769 \\
%
\hline
\end{tabular}
\end{center}
\end{table}

 From the variational principle
${\delta\Omega\over\delta\phi_i({\bf r})}=0$
($\phi_i=\sigma,R_0,\omega_0,V_{\rm Coul}$) and
${\delta\Omega\over\delta\rho_a({\bf r})}=0$ ($a=n,p,e$),
we get the coupled equations of motion for the mean-fields and the Coulomb
potential, 
\begin{eqnarray}
\nabla^2\sigma({\bf r}) &=& m_\sigma^2\sigma({\bf r}) +{dU\over d\sigma}
 -g_{\sigma N}(\rho_n^{(s)}({\bf r})+\rho_p^{(s)}({\bf r}))
    , \label{sigm}\\
\nabla^2\omega_0({\bf r}) &=& m_\omega^2\omega_0({\bf r}) -g_{\omega N}(\rho_p({\bf r})+\rho_n({\bf r}))
    , \label{omeg}\\
\nabla^2R_0({\bf r}) &=& m_\rho^2R_0({\bf r}) -g_{\rho N}(\rho_p({\bf r})-\rho_n({\bf r}))
    ,\\
\nabla^2{V_{\rm Coul}({\bf r})} &=& 4\pi e^2{\rho_{\rm ch}({\bf r})},  \label{puas} 
\end{eqnarray}
with the scalar densities $\rho_a^{(s)}({\bf r}), a=n,p$, and the charge density, ${\rho_{\rm ch}({\bf r})}={\rho_p({\bf
r})}+{\rho_e({\bf r})}$. Equations
of motion for fermions yield the standard relations between the densities
and chemical potentials,
\begin{eqnarray}
\mu_n &=& 
     \sqrt{k_{{\rm F},n}^2({\bf r})+{m_N^*({\bf r})}^2}
 + g_{\omega N}\omega_0({\bf r})-g_{\rho N}R_0({\bf r}),
  \label{eq:cpotB}\\
\mu_p &=&\mu_n-\mu_e 
   =  \sqrt{k_{{\rm F},p}^2({\bf r})+{m_N^*({\bf r})}^2}
   + g_{\omega N}\omega_0({\bf r})+
  g_{\rho N}R_0({\bf r})-{V_{\rm Coul}({\bf r})}\label{eq:cpotBp}, \\
{\rho_e({\bf r})}&=&-(\mu_e-{V_{\rm Coul}({\bf r})})^3/3\pi^2,
  \label{eq:rhoe}
\end{eqnarray}
where we have assumed the chemical equilibrium among nucleons and electrons.
The baryon-number chemical potential $\mu_B$ equals to $\mu_n$ 
and the charge chemical potential $\mu_Q$ to $\mu_e$.
Note that first, the Poisson equation for the Coulomb field (\ref{puas}) is a highly nonlinear
equation in
$V_{\rm Coul}({\bf r})$, since  $\rho_{\rm ch}({\bf r})$ in r.h.s. includes it in a
complicated way.  Secondly, the Coulomb potential always enters 
equations through the gauge invariant combinations $\mu_e-V_{\rm
Coul}({\bf r})$ and $\mu_p +V_{\rm
Coul}({\bf r})$. Thirdly, solutions of these equations of motion attain  
the kinematical equilibrium, i.e. equal pressure in each spatial point,
which is one of the Gibbs conditions.


\subsection{Numerical procedure}

To solve the above coupled equations numerically,
we use the Wigner-Seitz cell approximation:
the whole space is divided into equivalent cells with a geometry.
The geometrical shape of the cell changes:
sphere in three dimensional (3D) calculation, cylinder in 2D and slab in 1D,
respectively.
Each cell is globally charge-neutral and all the  physical quantities 
in a cell are smoothly connected to those of the next cell 
with zero gradients at the boundary.
Every point inside the cell is represented by the grid points (number of grids $N_{\rm grid}\approx 100$) and
the differential equations for fields are solved by the relaxation method
under the constraints of the given baryon number and the global charge neutrality.

To illustrate how to solve equations of motion for mean-fields,
let us consider, for simplicity, two fields $f_1(r)$, $f_2(r)$ 
and their coupled Poisson-like equations under 3D calculation,
\begin{eqnarray}
\nabla^2f_1(r)&=&{m_1}^2f_1(r)+W_1[f_1,f_2],\nonumber\\
\nabla^2f_2(r)&=&{m_2}^2f_2(r)+W_2[f_1,f_2],
\label{eq:poisson}
\end{eqnarray}
where $W_i(i=1,2)$ are functions of the fields $f_1$ and $f_2$.
Introducing a relaxation ``time'' $t$ we solve
\begin{eqnarray}
{df_i(r)\over dt} &=& c_i\left(\nabla^2f_i(r)-{m_i}^2f_i(r)-W_i\right).
\label{eq:solvemeson}
\end{eqnarray}
If coefficients $c_i$ are appropriately chosen, the above $f_i$ will
converge in time and we get the solution of Eq.~(\ref{eq:poisson}).

Baryon densities are solved 
with the help of the ``local chemical potentials'' $\mu_a(r)$ ($a=n,p$),
being different from the above introduced constant  chemical potentials.
Assuming 
$\mu_n(r)$ 
being an increasing function of the neutron density 
$\rho_n(r)$ in Eq.~(\ref{eq:cpotB}),
the relaxation equation for the neutron density,
\begin{eqnarray}
{d\rho_n(r)\over dt}&=&c_n(r)\; \rho_n(r)\nabla^2\mu_n(r),
\label{eq:solvebaryon}
\end{eqnarray}
is solved to equalize the local chemical potential $\mu_n(r)$ in each point.
The coefficient $c_n(r)$ is chosen to conserve the total neutron number.
The proton  density $\rho_p(r)$  is adjusted in the same way. 
When we impose the beta-equilibrium condition, proton and neutron densities
are adjusted to achieve $\mu_n(r)=\mu_p(r)+\mu_e(r)$.
Finally we get densities $\rho_n(r)$ and $\rho_p(r)$ 
leading 
the constant chemical potentials $\mu_n(r)=\mu_n$ and $\mu_p(r)=\mu_p$.
Nevertheless, the basic idea is to obtain constant chemical potentials,
$\mu_a(r)=\mu_a (a=n,p)$ at the convergence.
There is an exception: when there are some regions where $\rho_a(r)=0$,
the local chemical potential $\mu_a(r)$ is larger
than the corresponding constant value in the regions with finite $\rho_a(r)$.

The electron density $\rho_e(r)$ is calculated directly from 
Eq.~(\ref{eq:rhoe}).
The value of $\mu_e$ is adjusted at any time step
to get global charge neutrality:
we decrease $\mu_e$ when total charge in a cell is positive and
increase when it is negative.

All the above relaxation procedures are performed simultaneously, which
means that our numerical procedure always respect the Gibbs conditions
for chemical potentials.

\section{Bulk properties of finite nuclei}\label{Bulk}

\begin{figure}
\begin{minipage}[c]{0.65\textwidth}
\begin{center}
  \includegraphics[width=.49\textwidth]{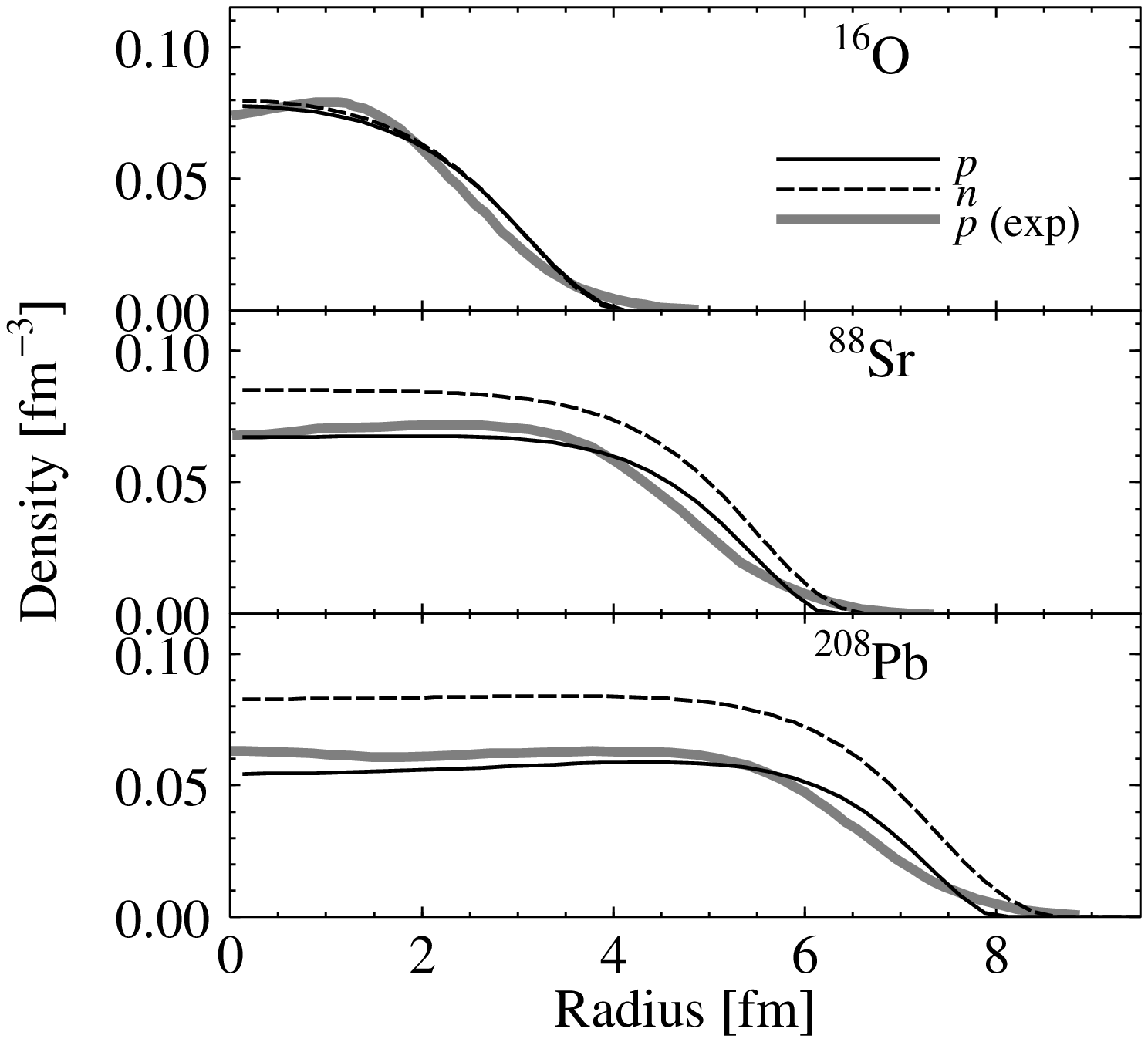}
  \includegraphics[width=.49\textwidth]{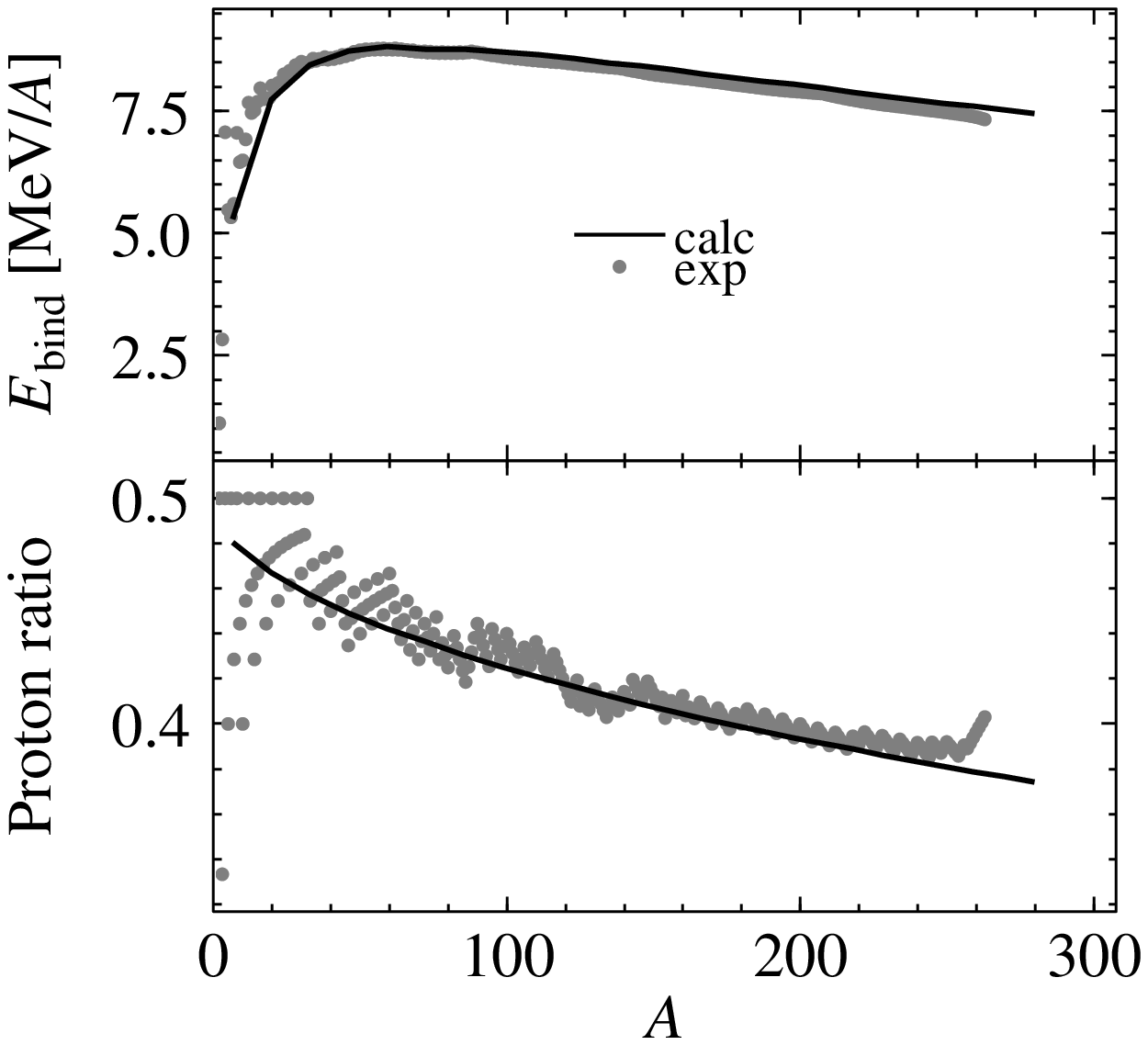}
\end{center}
\end{minipage}
\hspace{\fill}
\begin{minipage}[c]{0.33\textwidth}
\caption{Left: the density profiles of typical nuclei.
The proton densities (solid curves) are compared with the experiment.
Right: binding energy per nucleon and the proton ratio
of finite nuclei.\vspace{5mm}
\label{finite}
}
\end{minipage}
\end{figure}

Before applying our framework to the problem of the nuclear pastas  
, we check how it works to describe the bulk properties
of finite nuclei.
In this calculation for simplicity we assume  the spherical shape of nucleus.
The electron density is set to be zero. Therefore the global charge neutrality 
condition is not imposed.

In Fig.~\ref{finite} (left panel) we show the density profiles of some typical nuclei.
One can see how well our framework may reproduce density 
profiles.
To get a better fit, especially near the surface,
we could include the derivative terms of the nucleon densities, as we have mentioned.
Fine structures seen in the empirical density profiles, 
which come from the shell effects (see, e.g., a proton density  dip 
at the center of a light $^{16}$O nucleus),
cannot be reproduced by the mean-field approach. 
The effect of
the rearrangement of the proton density can be seen in heavy nuclei; protons repel each
other that enhances their concentration near the surface of the heavy nuclei. This
effect is analogous to the charge screening effect in a sense that the proton
distribution is now changed not on the scale of  the radius of the nucleus, as
for bare Coulomb field, but
on another length scale, that we  will call the proton Debye screening
length, see Eq.~(\ref{Deb}) below. 
It has important consequences for the pasta structures since typically the
proton Debye
length is less than the droplet size.
The stable value of the proton fraction $Y_p =Z/A$ ($Z$ and $A$ are proton 
and total baryon numbers, respectively)
is obtained by imposing the beta equilibrium condition $\mu_n=\mu_p$ 
for a given baryon number.
Figure \ref{finite} (right panel) shows the baryon number dependencies of the
binding energy per baryon and the proton ratio.
We can see that the bulk properties of finite nuclei 
(density, binding energy, and proton to baryon ratio) are satisfactorily 
reproduced for our present purpose.

\begin{figure}
\includegraphics[height=.38\textheight]{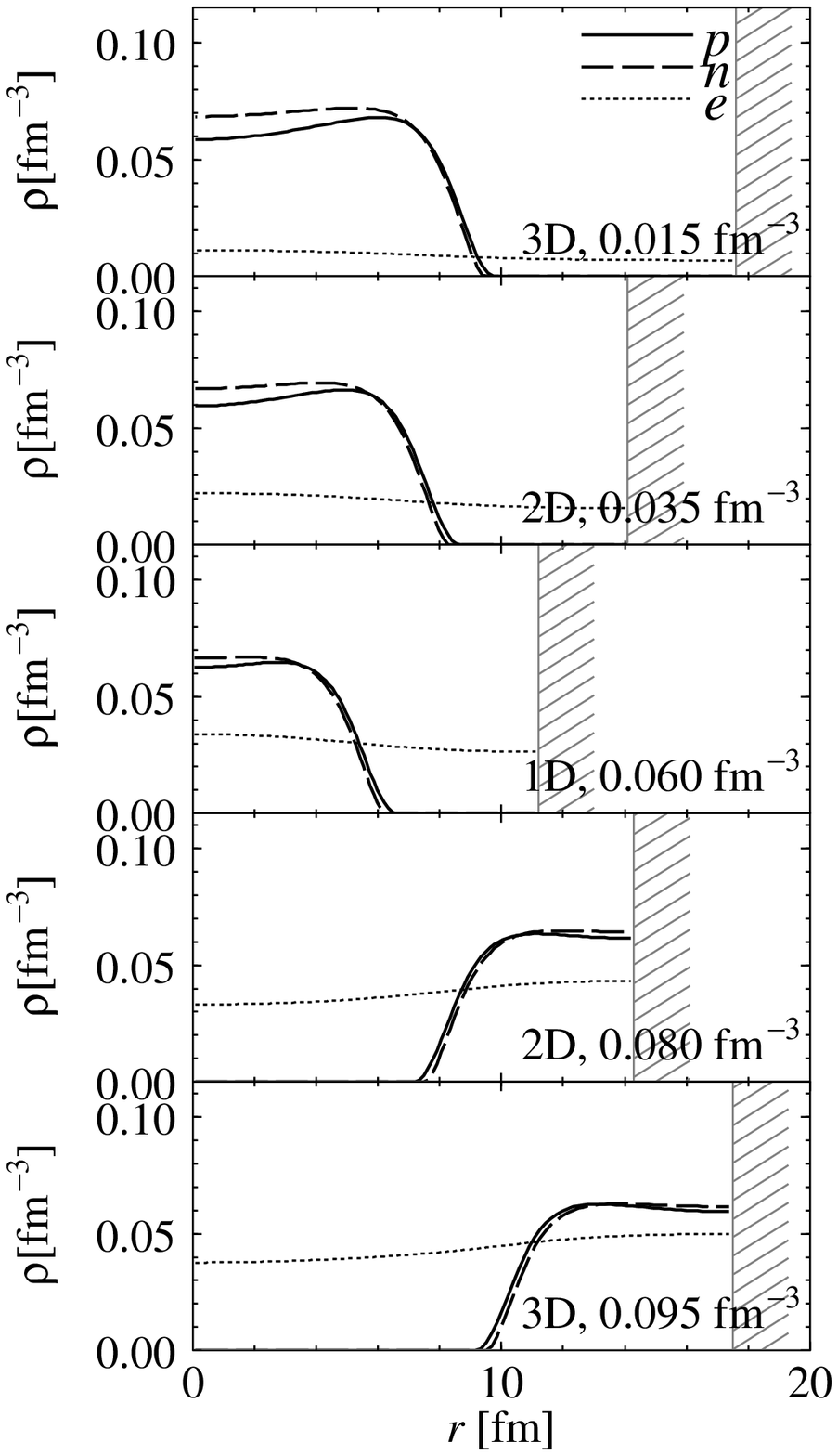}
\includegraphics[height=.38\textheight]{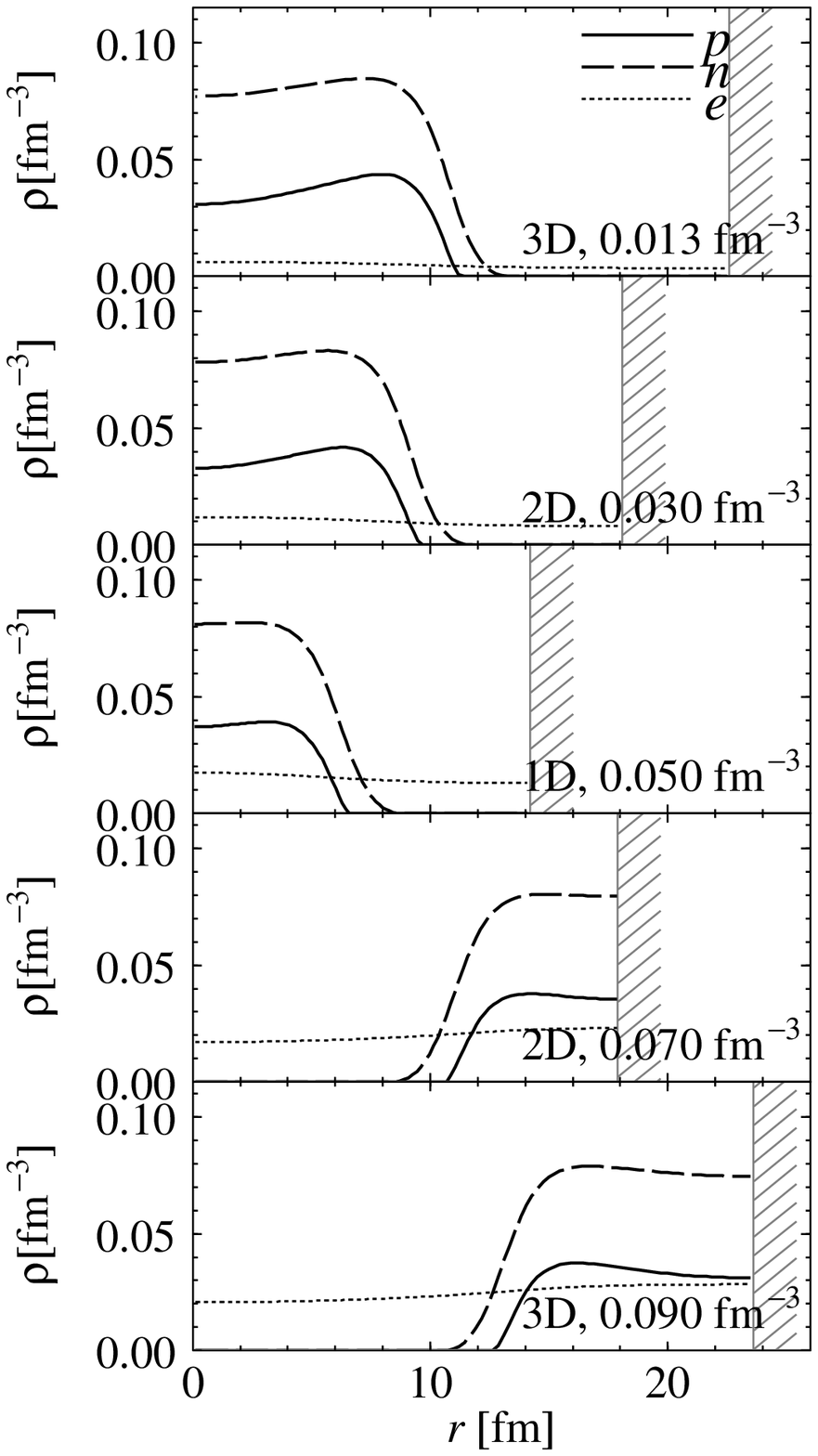}
\includegraphics[height=.38\textheight]{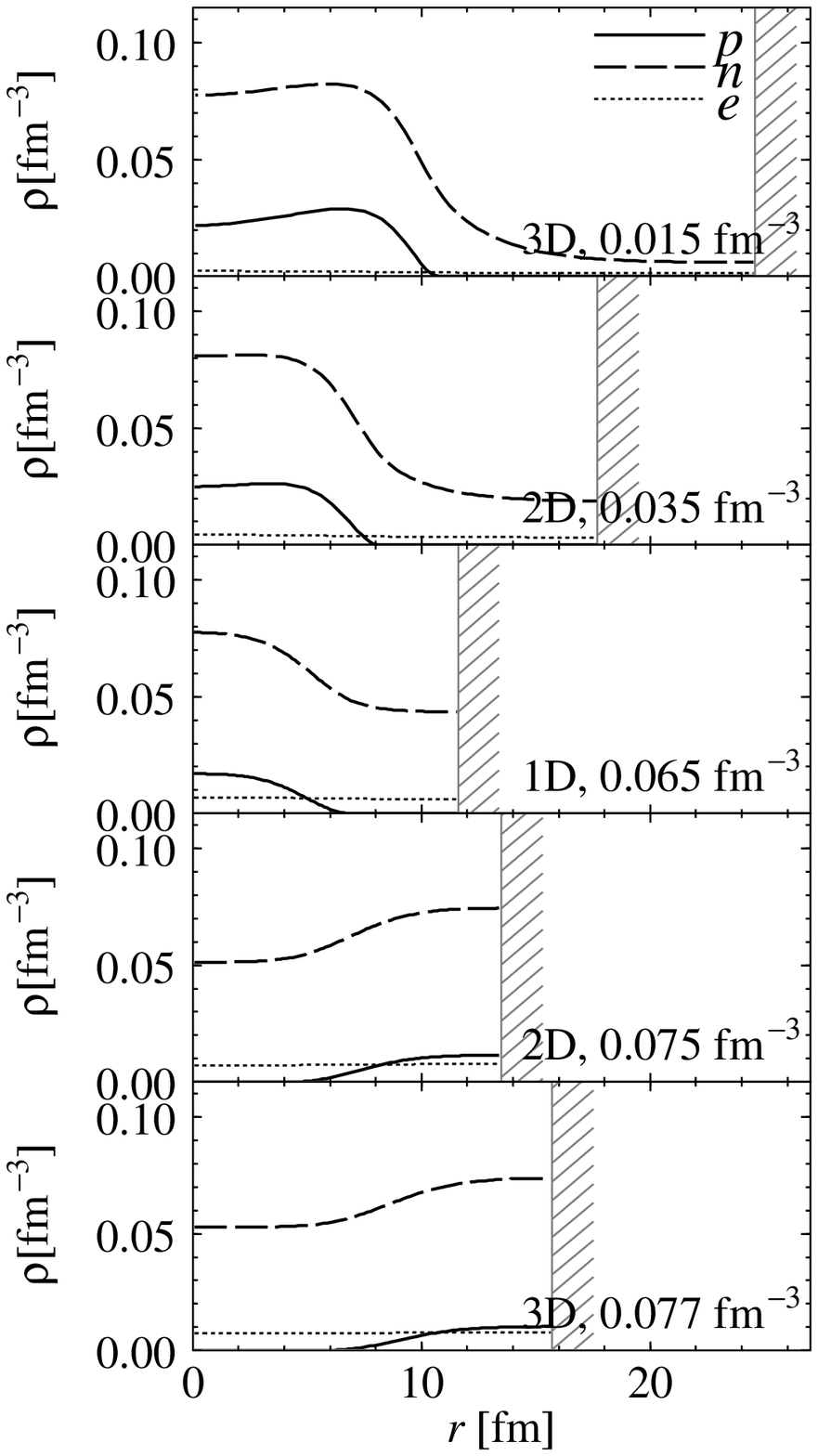}
\caption{
Examples of the density profiles in the cell for symmetric nuclear
 matter with $Y_p$=0.5 (left) and for asymmetric matter
 with $Y_p=0.3$ (center) and 0.1 (right).
}
\label{proffixfull}
\end{figure}

Note that we must use a slightly smaller value of the  sigma mass
i.e.\ 400 MeV, 
than that one usually  uses 
to get an appropriate fit.
If we used a popular value $m_\sigma\approx 500$ MeV,
finite nuclei would be overbound by about 3 MeV/$A$.
The actual value of the  sigma mass (as well as the omega and rho masses)
has little relevance for the case of infinite nuclear  matter,
since it enters the thermodynamic potential only in the combination
$\widetilde{C}_{\sigma}=g_{\sigma N}/m_{\sigma}$.
However meson masses  are important characteristics of finite
nuclei and of
other non-uniform nucleon systems, like those in pastas.
The effective meson mass characterizes the typical scale for the spatial change of the
meson field 
and consequently it affects, e.g., the surface property of the given nucleon structure
and it influences the  value of the effective surface tension.

\section{Non-uniform structures in nuclear matter}\label{Non-uniform}

\subsection{Nucleon  matter at fixed proton fraction}

\begin{figure*}[t]
\includegraphics[height=.28\textheight]{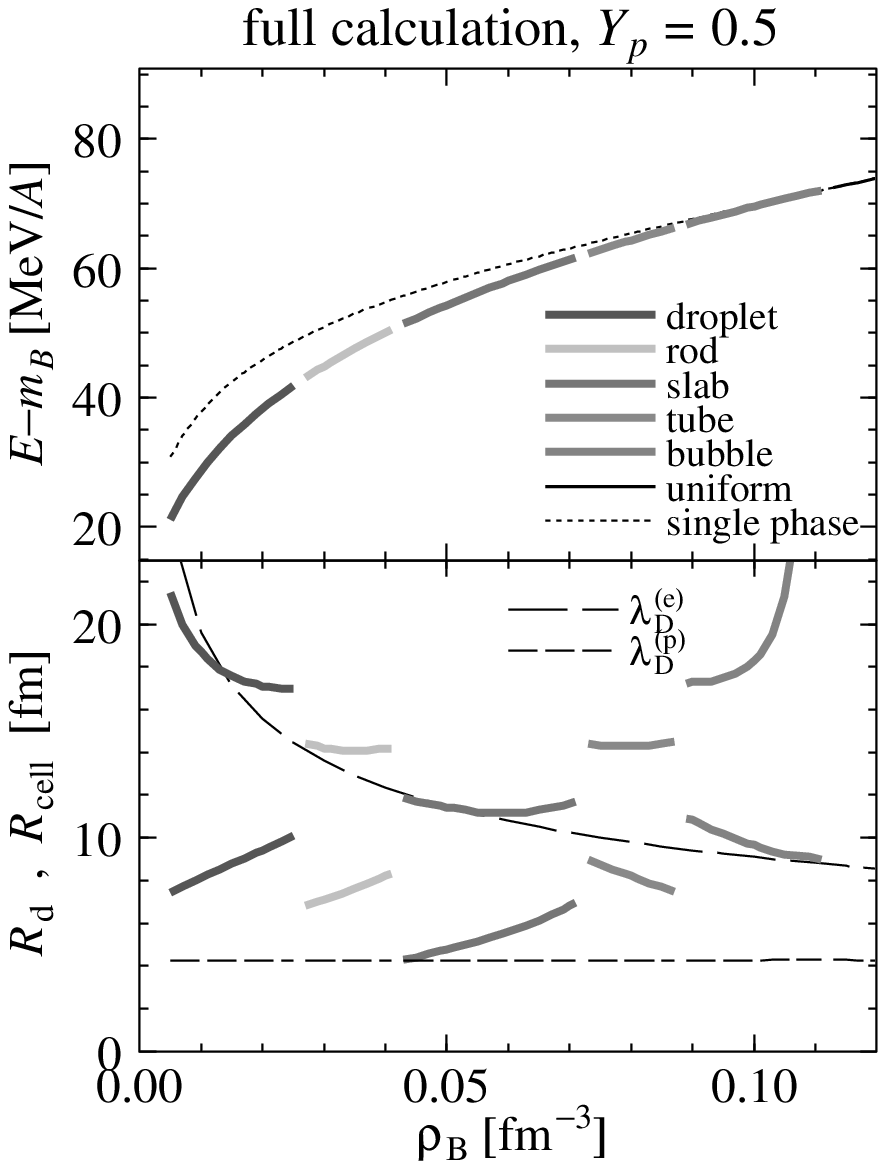}
\includegraphics[height=.28\textheight]{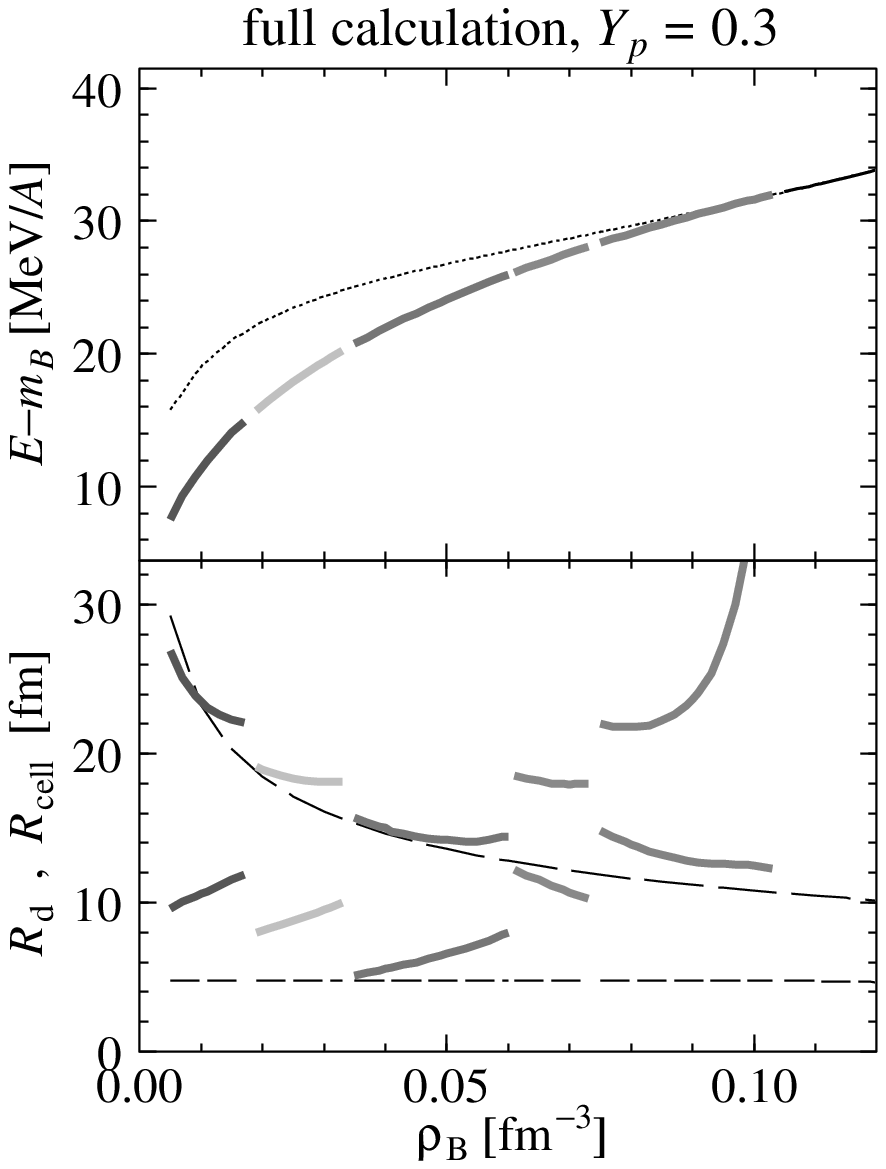}
\includegraphics[height=.28\textheight]{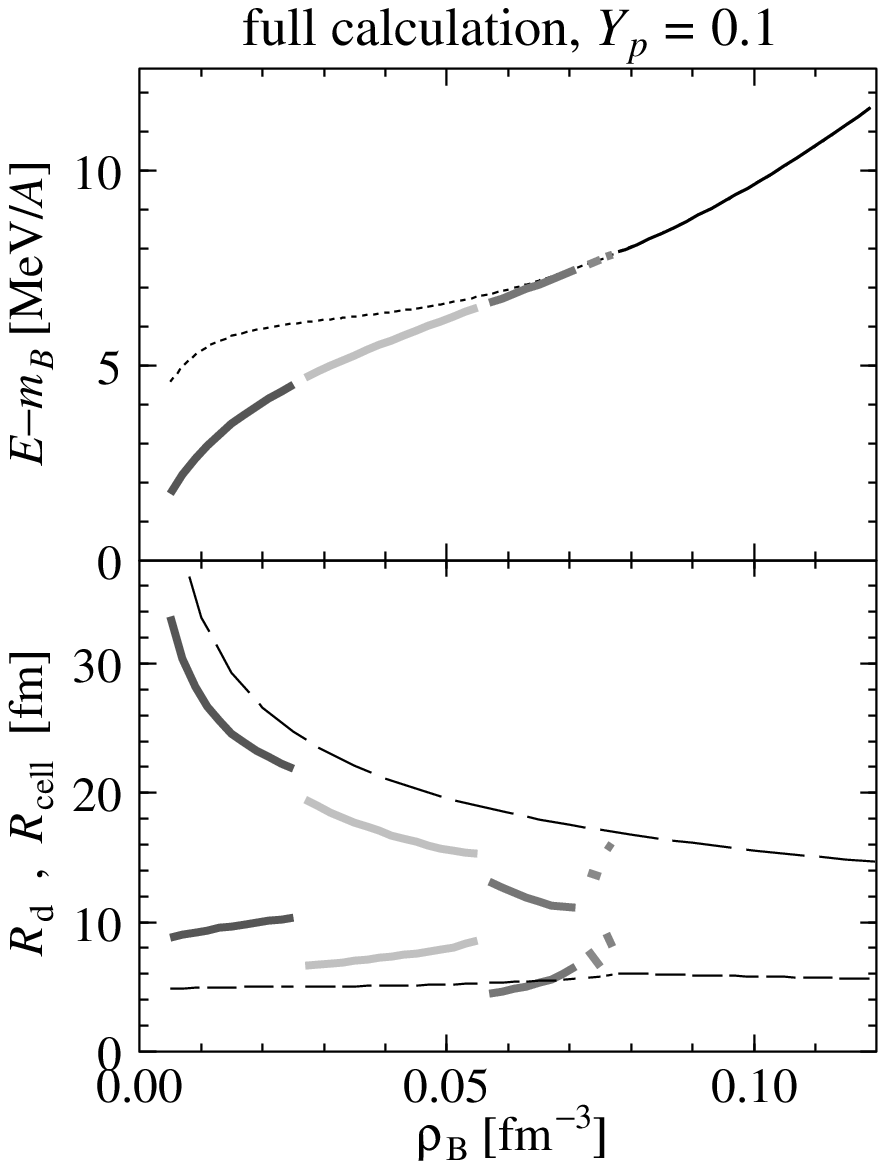}
\caption{
Binding energies per nucleon and the cell size for symmetric nuclear
 matter with $Y_p$=0.5 (left), and for asymmetric matter with
 $Y_p=0.3$ (middle) and 0.1 (right).
}
\label{eosfixfull}
\end{figure*}

First, we focus on the discussion of the behavior of the nuclear matter at a fixed proton fraction $Y_p$.
Particularly, we explore  proton fractions $Y_p=0.1$, 0.3, and 0.5.
The cases $Y_p=0.3$ -- 0.5 might be relevant for the supernova matter and 
for newly
born neutron stars.
\footnote{Note that we have no relation among nucleon chemical potentials
and electron and three chemical potentials become independent in this case, 
different from the usual $\beta$ stable matter. Accordingly we 
impose one more condition on the system as a Gibbs condition.}
Figure \ref{proffixfull} shows some density profiles inside the Wigner-Seitz
cells as functions of the radial distance from the center of the cell.
The geometrical dimension of the cell  is denoted as ``3D'' (three
dimensional), etc. 
The cell boundary is indicated by the hatch.
 From the top to the bottom the configuration changes like  
droplet (3D), rod (2D), slab (1D), tube (2D), and bubble (3D).
Thus the nuclear pastas are clearly manifested.
For the lowest $Y_p$ case ($Y_p =0.1$), the neutron density is finite at any point:
the space is filled by dripped neutrons. For a higher  $Y_p$, the neutron density
drops to zero outside the structure. The proton density always drops to
zero outside the nuclear lump.
The charge screening effects are pronounced, especially due to protons. 
Protons repel each
other and  
thereby the proton density profile substantially deviates from the
step-function: the proton
number is enhanced near the surface of the nuclear lump.
The electron density also becomes non-uniform by the rearrangement effect.
This non-uniformity of the electron distribution is
more pronounced for a higher $Y_p$ and a higher density.

EOS of the nuclear pastas  
is shown as a function of the averaged density in Fig.~\ref{eosfixfull} 
(upper panels).
The energy $E-m_B$ includes the kinetic energy of electrons,
which makes the total pressure positive.
The lowest-energy configuration is selected among various geometrical
structures for a given averaged density.
The most favorable configuration changes from the droplet to rod, slab, tube, bubble,
and to the uniform one (the dotted thin curve) with increase of the density. 
We can see that the appearance of the nuclear pastas  results in
a softening of the EOS:
the energy per baryon gets lower up to about 15 MeV$/A$ 
compared to the uniform case.
The lower panels in Fig.~\ref{eosfixfull} show the cell sizes $R_{\rm cell}$ and 
structure sizes $R_{\rm d}$ as functions of the averaged density.
The size $R_{\rm d}$ is 
defined here by way of a density fluctuation as
\begin{eqnarray}
R_{\rm d}&=&
\cases{
\displaystyle
R_{\rm cell} { \langle\rho_p\rangle^2 \over \langle\rho_p^2\rangle },\ \ \ \hbox{(for droplet, rod, and slab)}\cr
\displaystyle
R_{\rm cell} \left(1-{\langle\rho_p\rangle^2 \over \langle\rho_p^2\rangle}\right),\ \ \ \hbox{(for tube and bubble)} }
\end{eqnarray}
where the bracket ``$\langle \rangle$'' indicates the average 
inside the cell.
Dashed curves show the Debye screening lengths of 
electron and proton calculated as
\begin{equation}\label{Deb}
\lambda^{(e)}_D=\left(-4\pi e^2{d\rho_e^{\rm av}\over d\mu_e}\right)^{-1/2},~~~
\lambda^{(p)}_D=\left(4\pi e^2{d\rho_p^{\rm av}\over d\mu_p}\right)^{-1/2},
\end{equation}
respectively, where
$\rho_p^{\rm av}$ is the proton density
averaged over the nuclear lump 
and
$\rho_e^{\rm av}$ is the electron density
averaged inside the cell.
%
Note that these quantities are obviously gauge invariant.
Numerically, the cell size $R_{\rm cell}$ for droplet, rod, and slab
configurations at $Y_p=0.5$ and 0.3 are shown to be close to
the Debye screening length of electron. 
For $Y_p=0.1$, in all cases  $R_{\rm cell}$ is substantially
smaller than $\lambda^{(e)}_D$ and  thereby the electron screening should be much weaker.
In all cases, except for bubbles (at $Y_p=0.5$ and 0.3), the structure size $R_{d}$ are
smaller than $\lambda^{(e)}_D$.
This means that the Debye screening effect of electrons inside these structures should not be
pronounced. For bubbles at $Y_p=0.5$ and 0.3, $\lambda_D^{(e)}$
is substantially smaller than the cell size and the electron screening should
be significant.
%
For $Y_p=0.5, 0.3, 0.1$ in all  cases (with the only exception $Y_p=0.1$ for
slabs), the value $\lambda_D^{(p)}$ is shorter than $R_{\rm d}$, which
means that the rearrangement of proton density is essential
for the structures of the nuclear pastas, as it is indeed seen from the Fig.~\ref{proffixfull}. 

\medskip
\centerline {
\hfill\vbox{\hsize15.5cm\hrule\hbox to 15.5cm{\vrule\hfill\vbox to 1.5cm{\hsize15cm
\medskip\noindent
{\bf Problem}: 
Evaluate the electron Debye screening length 
in the case of $\rho_e^{\rm av}=-0.5\rho_0$, by using 
the above expression Eq.\ (\ref{Deb}) for massless electrons.
\vfill}\hfill\vrule}\hrule}\hfill}

\begin{wrapfigure}[22]{r}{0.45\textwidth}
\begin{center}
\vspace*{-5mm}
\includegraphics[height=.39\textheight]{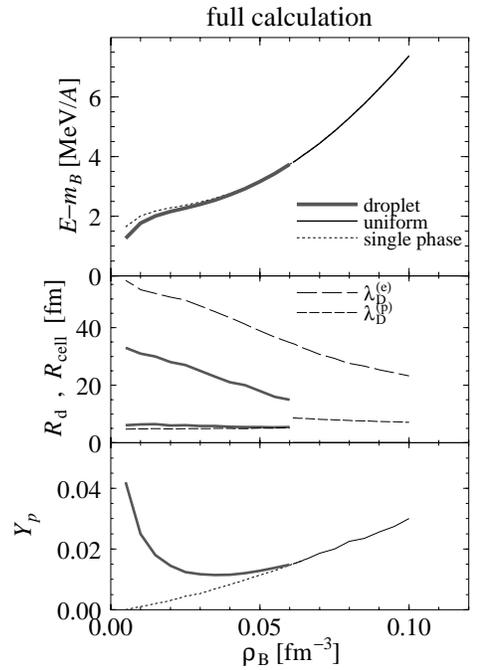}
\caption{
Binding energy (top), structure size (middle), and proton ratio (bottom)
in the cell for nuclear matter in beta equilibrium.
}
\label{eosbeta}
\end{center}
\end{wrapfigure}

Using the baryon density and the structure size from  Fig.~\ref{eosfixfull},
one may estimate the atomic number of the structure.
In the case of droplets and for $Y_p=0.5$ the atomic number of the droplet is
$\simeq 25$ at lower 
density limit
and $\simeq 65$ at the maximum density of the droplet phase
$\rho_{B,d}^{(\rm max)} \simeq 0.025\ {\rm fm}^{-3}$.

\subsection{Nuclear matter in beta equilibrium}

%

Next, we 
explore the nuclear pastas in beta equilibrium. 
The droplet structure appears, which is quite similar to the 
above considered case of the fixed proton ratio $Y_p=0.1$.
The apparently different feature in this case is that
only the droplet configuration appears as a non-uniform structure.
It should be noticed, however, that the presence or absence of the
each pasta structure may 
sensitively depend on the choice of the effective interaction.

In Fig.~\ref{eosbeta} we plot 
EOS (top), the structure size (middle), and the proton ratio (bottom).
The difference between
EOS of uniform matter and that of non-uniform one is small, while 
the proton ratio is significantly affected by the presence of the pasta at lower densities.
The droplet radius and the cell radius in the middle panel of Fig.~\ref{eosbeta}
are always smaller than the electron Debye length $\lambda_D^{(e)}$, and
thereby the effect of the electron charge screening is small. 
On the other hand,  the proton Debye
length $\lambda_D^{(p)}$ is comparable with the droplet radius at all
densities, which demonstrates the relevance of the proton screening.


\section{Charge screening effect in nuclear pastas}\label{Charge}

%

\begin{figure}
\includegraphics[height=.28\textheight]{fig/eos-full05-5BW.ps}
\includegraphics[height=.28\textheight]{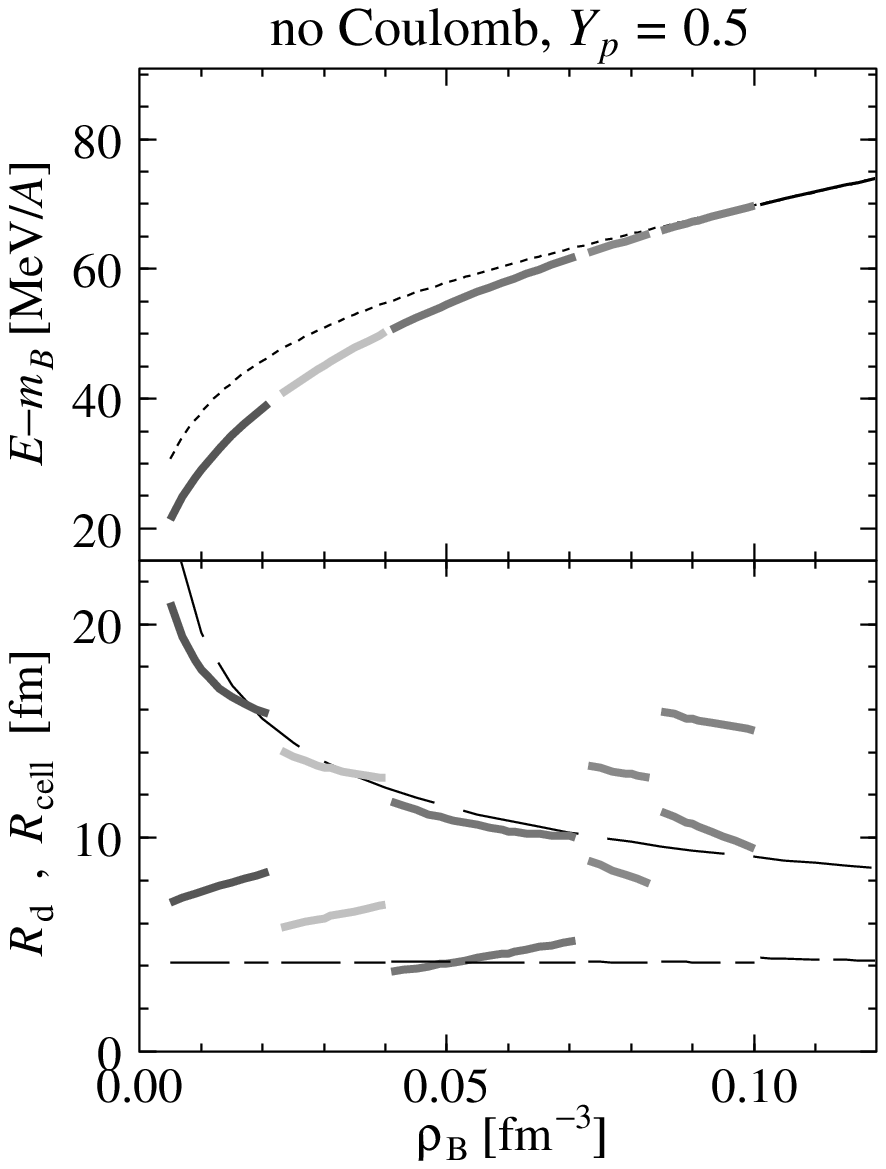}
\includegraphics[height=.28\textheight]{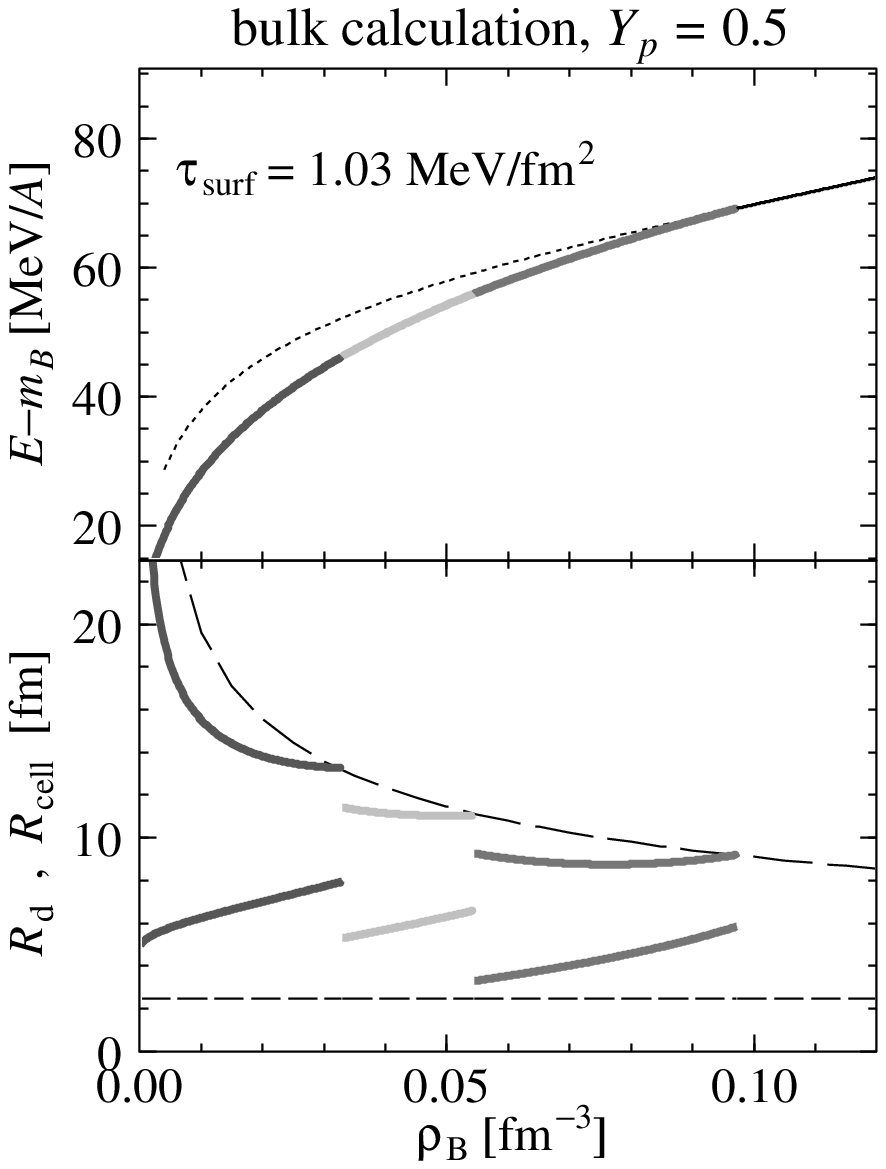}
\caption{
Comparison of the density profiles for
different treatments of the Coulomb interaction.
 From the left: ``full'' calculation, 
``no Coulomb'' calculation,
and ``bulk'' calculation.
The proton ratio for all cases is $Y_p$=0.5.
}
\label{eoscompare}
\end{figure}


In this section we explore the effect of the charge screening on the
nuclear pastas.
Here we focus on the matter with fixed proton ratio $Y_p=0.5$ since 
the Coulomb effects should be most pronounced in this case.
We compare three kinds of calculations with different treatments
of the Coulomb interaction.
One is the ``full calculation'' which we have presented above.
The second calculation (``no Coulomb'') is performed by totally discarding 
the Coulomb potential $V_{\rm Coul}$ in equations of motion. 
After getting the density profiles, the
Coulomb energy, being evaluated using
charge densities thus determined, is added to the total energy. 
Note that this calculation is similar but not the same as the bulk calculations: in
the latter a sharp boundary and the surface tension are introduced, and 
the particle density profile is assumed to be
constant in each phase.
The third  is  
the bulk calculation, in which we use
a completely uniform electron density,
the baryon density distribution of a step function,
and the surface tension introduced by hand.
To determine the geometrical structure we have used
the equilibrium condition of baryon chemical potential and the 
pressure calculated the present RMF model,
and the relation between surface energy $E_{\rm surf}$ and
the Coulomb energy $E_{\rm Coul}$ \cite{Rav83} as
\begin{eqnarray}
E_{\rm surf}&=&2E_{\rm Coul}.
\end{eqnarray}
We  have used the surface tension parameter $\tau=1.03$ MeV/fm$^2$ to
fit the liquid-drop binding energies of finite nuclei.
%


EOS as a whole (upper panels in Fig.~\ref{eoscompare})
shows almost no dependence on the calculation.
This agrees with a general
statement that the variational functional is
always less sensitive to  the choice of the  trial functions than the quantities
depending on  these  trial functions linearly.

Nevertheless,
the density region of pasta structures and the
sizes of the structures (lower panels in Fig.~\ref{eoscompare})
especially for tube and bubbles are different.
In fact we don't observe tube and  bubble configurations in the
present bulk calculation.
In Ref.~\cite{Rav83} and others, however, they have reported the appearance
of full pasta structures by the bulk calculation.
The most important origin of this discrepancy may 
come from the difference in the nuclear EOS used in the calculation.
Also the surface tension is crucial; if we use smaller value of $\tau$,
e.g.\ 0.3 MeV/fm$^2$, we observe full pasta structures appear 
in wider density range.
If we take $\tau=0$ the mixed phase spreads from zero 
to the saturation density $\rho_0$ without any specific geometry.
From the bulk calculation we see that the surface tension 
plays a crucial role in the appearance of pasta structures.

Comparing the ``no Coulomb'' calculation with ``full calculation'',
we don't see large difference.
However, precisely looking, the density region of pasta in ``full calculation''
is slightly larger.
The density dependence of the cell size is also different.
In the case of ``no Coulomb'' all the pieces of lines for $R_{\rm cell}$ 
are monotonously decreasing.
The behavior of $R_{\rm cell}$ in the full calculation is
rather similar to the bulk calculation.

\section{Kaon condensation in high-density matter}

\begin{figure}
\begin{minipage}{0.48\textwidth}
\includegraphics[height=.25\textheight]{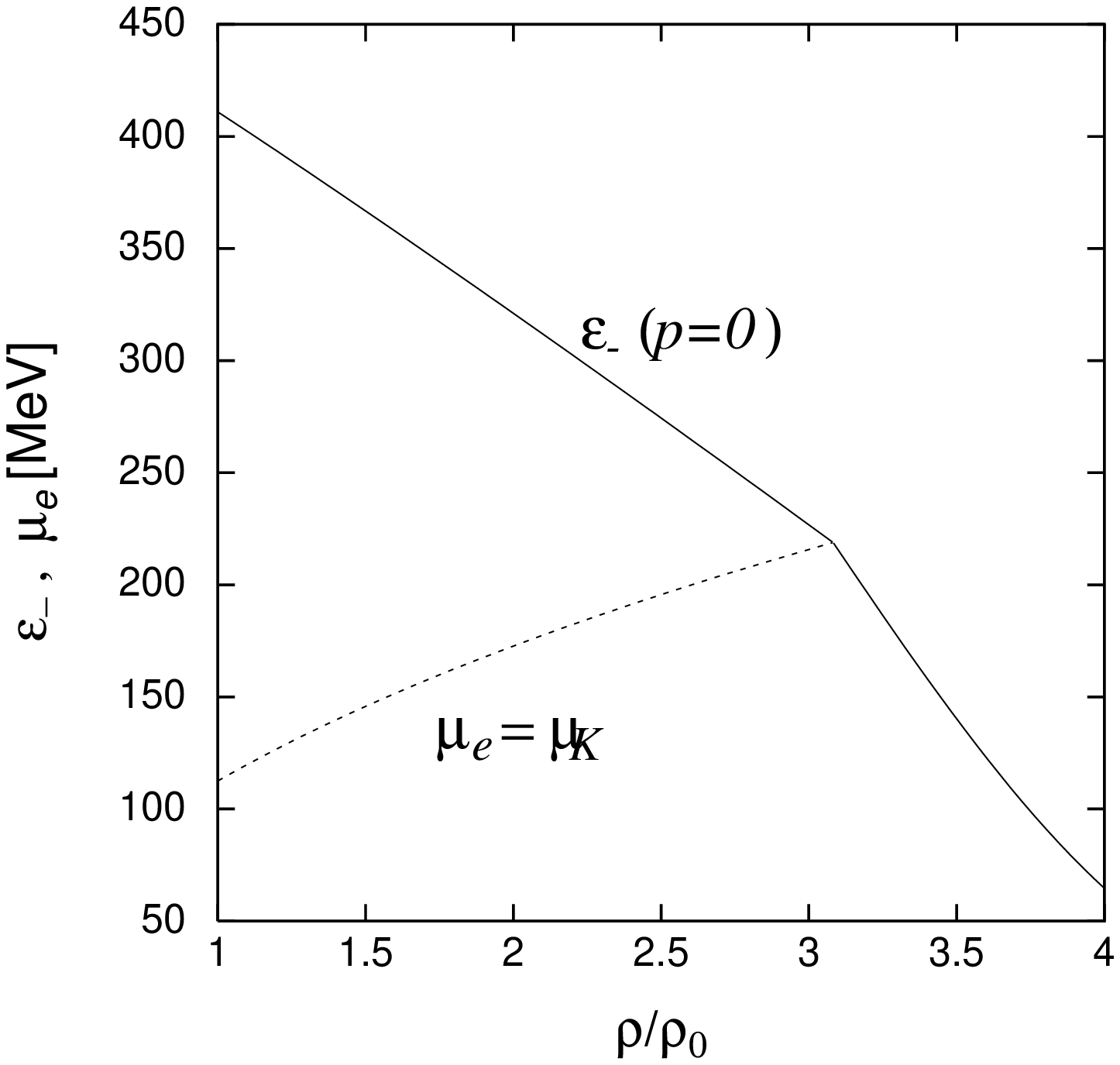}
\caption{Onset mechanism of kaon condensation. 
}
\label{OnsetKaon}
\end{minipage}
\hspace{\fill}
\begin{minipage}{0.48\textwidth}
\includegraphics[height=.25\textheight]{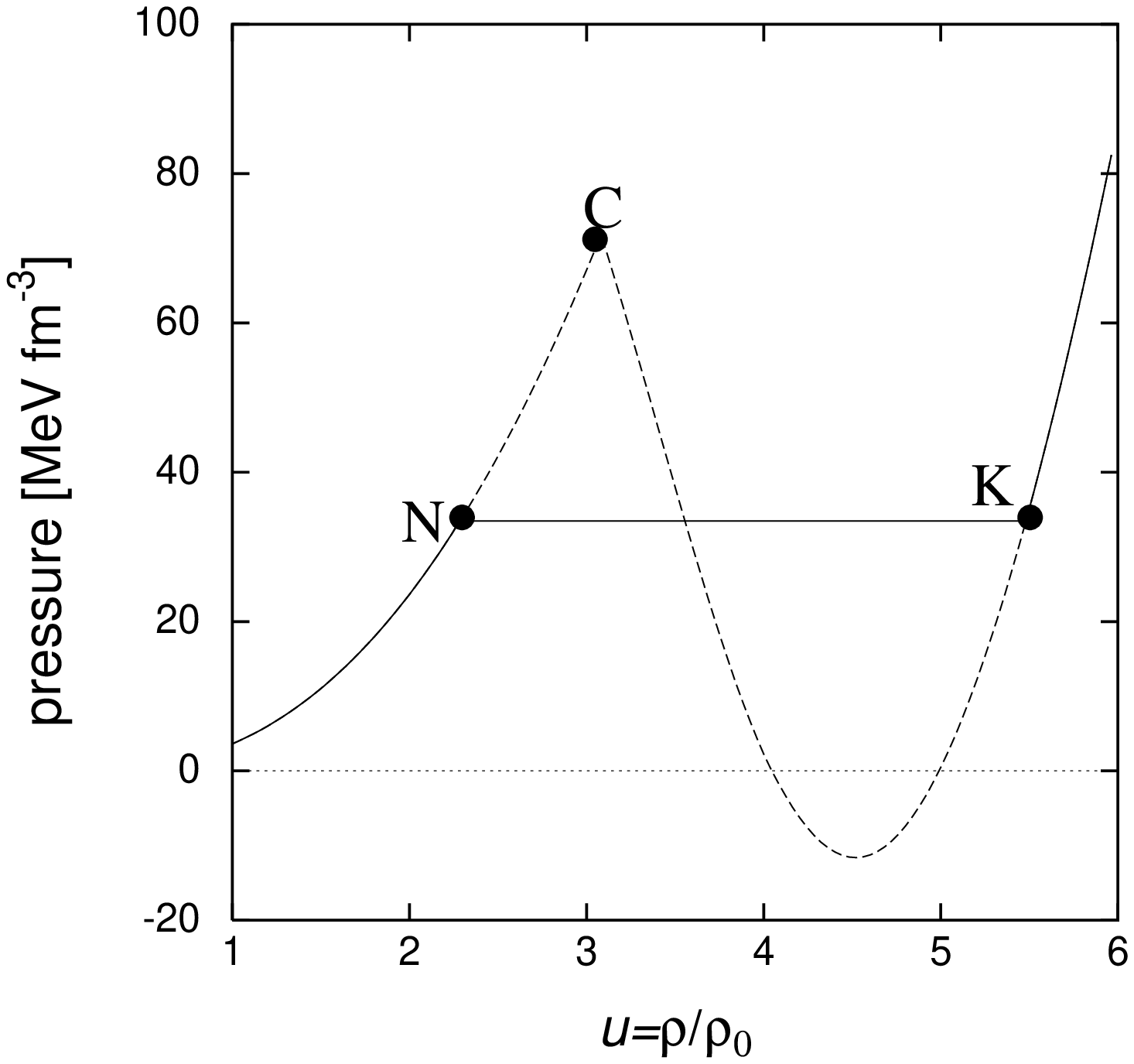}
\caption{Original EOS and the Maxwell construction.
}
\label{EOSMaxwell}
\end{minipage}
\end{figure}

Next we explore the high-density nuclear matter in beta-equilibrium, which is 
expected to exist in the inner core of neutron stars. Kaons are the lightest mesons with 
strangeness, and their effective energy is much reduced by the 
kaon-nucleon interaction in nuclear medium.
For low-energy kaons the $s$-wave interaction is dominant and attractive in 
the $I=1$ channel, so that negatively charged kaons appear in the neutron-rich 
matter once the process $n\rightarrow p+K^-$ becomes energetically allowed. 
Since kaons are bosons, it causes the Bose-Einstein condensation at momentum ${\bf p}=0$ \cite{kn86}. 
The threshold condition then reads
\begin{equation}
\epsilon_-({\bf p}=0)=\mu_n-\mu_p=\mu_e,
\end{equation}
which means the kaon distribution function diverges at ${\bf p}=0$ (Fig.~\ref{OnsetKaon}).

If kaon condensation occurs in nuclear matter, it has many implications on 
compact stars; softening of EOS may give the possibility of the delayed 
collapse of a supernova to the low-mass black hole, and the nucleon Urca 
process under background kaons may give a fast cooling mechanism of neutron 
stars \cite{t95,bb94,bkpp88,t88,pb90}. 

Since many studies have shown that kaon condensation is of the first order, we 
must carefully treat the phase change (Fig.~\ref{EOSMaxwell}).  In the following we
discuss the kaon mixed phase in a similar way to nuclear pastas. There
have been some studies about the mixed phase in the kaon condensation
\cite{GS99,CG00,CGS00,PREPL00,MYTT,RB,NR00,maruKaon}. In ref.\cite{NR00} the charge screening effect has been also
studied, but all the equations of motion have not been solved self-consistently.


\begin{table}
\caption{
Additional parameters used in RMF with kaon. The kaon optical $U_K$ is
 defined by $U_K = g_{\sigma K}\sigma+g_{\omega K}\Omega_0$.
}
\begin{center}
\begin{tabular}{ccccc}
\hline
 $f_K (\approx f_\pi)$ [MeV] &
 $m_K$ [MeV]&
 $g_{\omega K}$ &
 $g_{\rho K}$ &
 $ U_K(\rho_0)$ [MeV]
\\ [1mm]
\hline\\
 93  &
 494  &
$g_{\omega N}/3$ &
$g_{\rho N}$ &
$-120$ -- $-130$ \\
\hline
\end{tabular}
\end{center}
\end{table}

To incorporate kaons into our calculation,
the thermodynamic potential of Eq.~(\ref{Omega-tot}) is modified as
\begin{eqnarray}
\Omega &=& \Omega_N+\Omega_M +\Omega_e +\Omega_K,\\
\Omega_K&=&\!\!\!\int d^3r\left[
  -{f_K^2\theta^2\over2}\left[-{m_K^*}^2+(\mu_K-V_{\rm Coul}+g_{\omega K}\omega_0
  +g_{\rho K}R_0)^2 \right]+{f_K^2(\nabla\theta)^2\over2} \right],
\end{eqnarray}
where $m_K^*=m_K-g_{\sigma K}\sigma$,\ \ $\mu_K=\mu_e,$\ \ and 
the kaon field $K=f_K\theta/\sqrt{2}$ ($f_K$ : Kaon decay constant).\footnote{%
We here consider a linearized $KN$ Lagrangian for simplicity,
which is not chiral-symmetric.}
The equations of motion are then similar to Eqs.~(5) - (11) given for
nuclear pastas except kaon contributions \cite{maruKaon}.
Additional parameters concerning kaons are presented in Table 2.

\begin{wrapfigure}[24]{r}{0.45\textwidth}
  \includegraphics[height=.42\textheight]{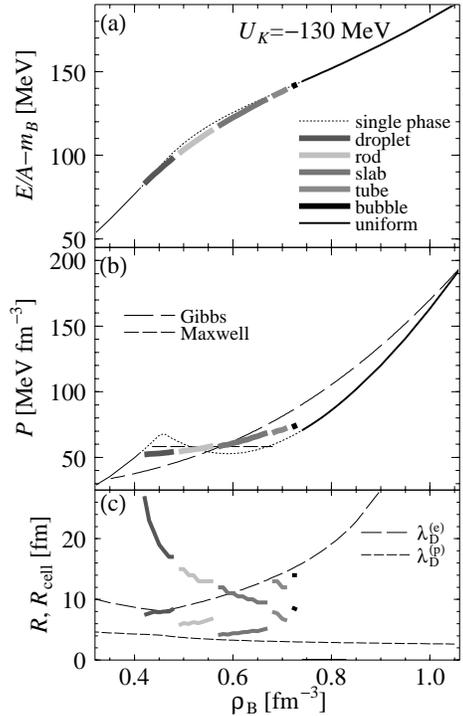}
\caption{
  Top: binding energy per nucleon of
  nuclear matter in  beta equilibrium.
  Middle: pressure.
  Bottom: structure size $R$ (thick curves below)
  and the cell size $R_{\rm cell}$ (thick curves above).
}\label{figKEOS}
\end{wrapfigure}

If Glendenning's claim were correct, the structured mixed phase
would develop in a broad density range from well below to
well above the critical density \cite{GS99,CG00,CGS00}.
In this  density interval the matter should exhibit the structure change
similar to the nuclear pastas \cite{maru1}:
the kaonic droplet, rod, slab, tube, bubble.
Actually we observe such structures (kaonic pastas) in our calculation.
%
In the top  (a) and the middle (b) panels of Fig.~\ref{figKEOS} we show EOS;
pieces of solid curves indicate the energetically most favored structures,
while the dotted curve EOS of the uniform matter.
One can see the softening of EOS by the appearance of kaonic pastas.
In the bottom panel (c) plotted are
the size of the kaonic lump or hole 
and the cell size.
The dashed lines 
show the Debye screening
lengths of the electron and the proton,
$\lambda_{D}^{(e)}$ and $\lambda_{D}^{(p)}$, respectively.
In most cases
$\lambda_{D}^{(e)}$ is less than the cell size $R_{\rm cell}$
but it is larger than the structure size $R$.
However the proton Debye length $\lambda_{D}^{(p)}$
is always shorter than $R_{\rm cell}$ and $R$.
When the minimal  value of the Debye length inside the structure
is shorter or of the order of $R$,
the charge screening effects  should be pronounced \cite{voskre}.

\section{Charge screening effect in the kaon mixed phase}\label{ChargeK}
\begin{figure}[t]
  \includegraphics[width=.45\textwidth]{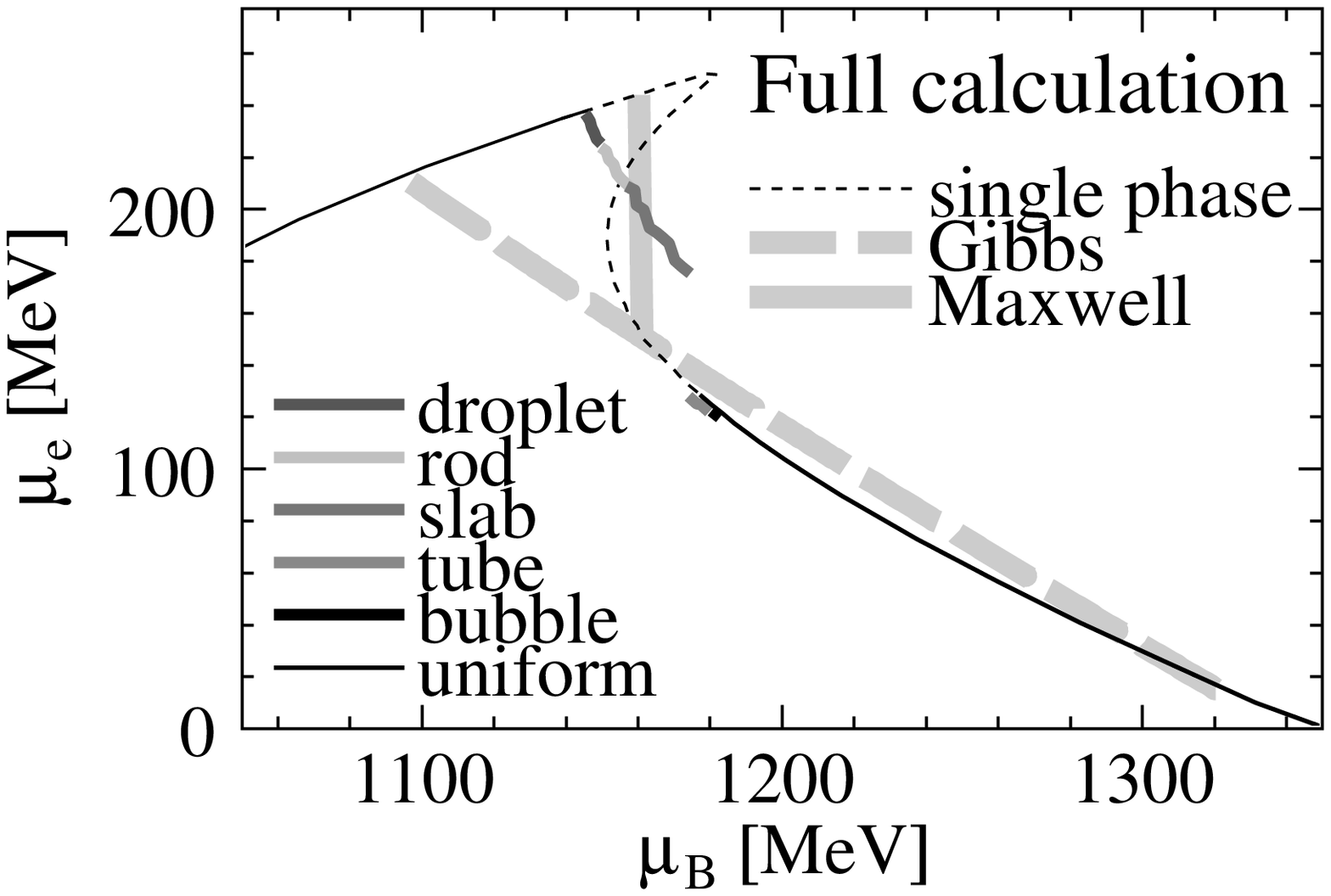}
  \includegraphics[width=.45\textwidth]{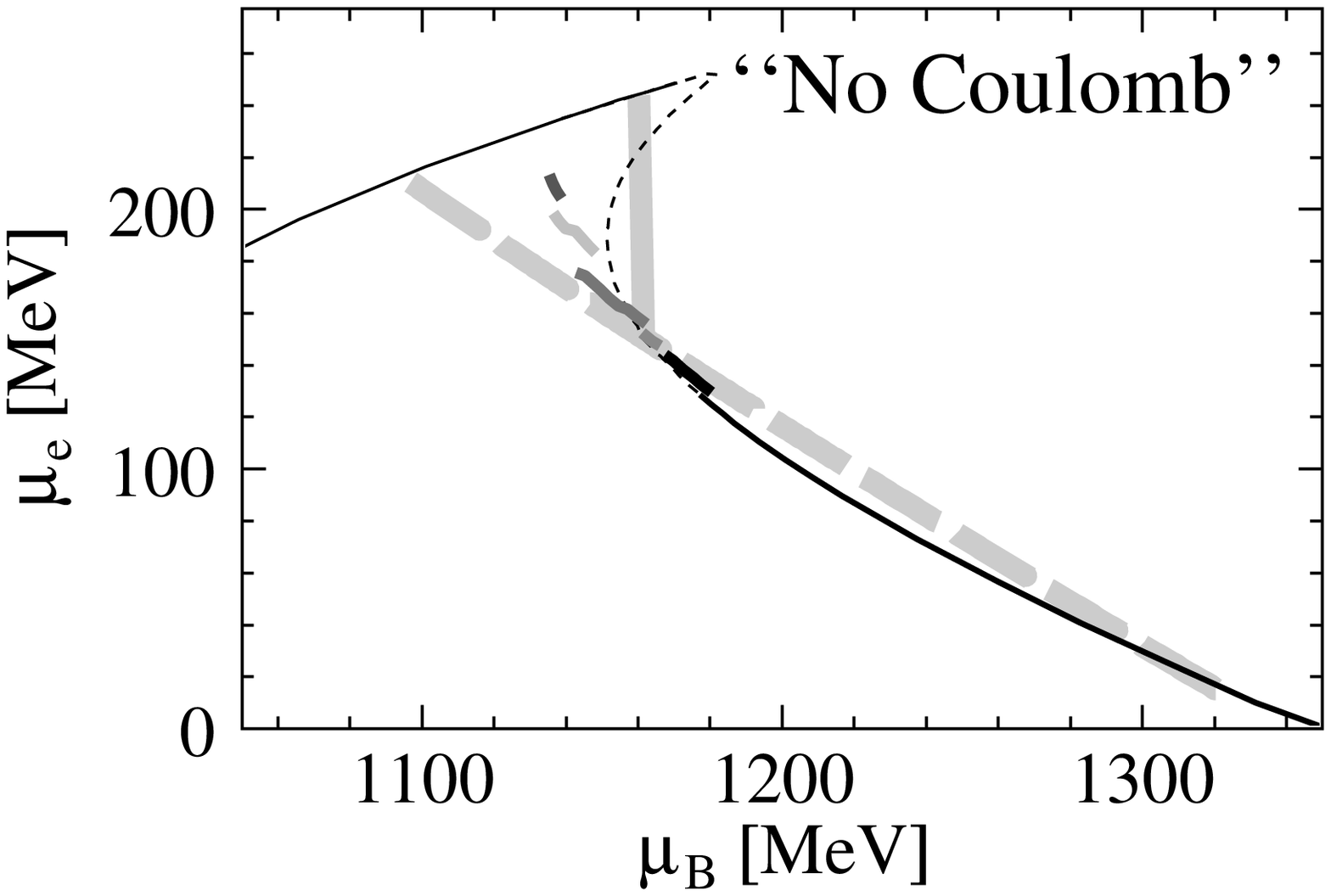}
  \caption
{Phase diagram in the $\mu_B$-$\mu_e$ plane,
where $\mu_B$ and $\mu_e$ denote
baryon- and charge-chemical potentials, respectively.
Left: full calculation.
Right: the electric potential is discarded
in determining the matter structure.
Curves calculated using  Gibbs conditions disregarding finite size effects
and the Coulomb interaction effect (``Gibbs'')
and that for the Maxwell construction (``Maxwell'') are also presented
 for comparison.
}\label{figKmumu}
\end{figure}

To demonstrate the charge screening effect on the kaonic mixed phase and
discuss differences among various treatments
,
we show pressure $P$ ($=-\Omega/V$) in the middle panel (b) of 
Fig.~\ref{figKEOS} and the phase diagrams in the $\mu_B$-$\mu_e$
plane  in Fig.~\ref{figKmumu}.
From Fig.~\ref{figKEOS} (b) we can clearly see that our results give the similar
pressure to the one given by MC, while the bulk calculation, where no
Coulomb interaction or surface energy is taken into account, gives a wide
density region for the mixed phase.

\begin{figure}
\begin{minipage}[c]{0.55\textwidth}
  \includegraphics[height=.30\textheight]{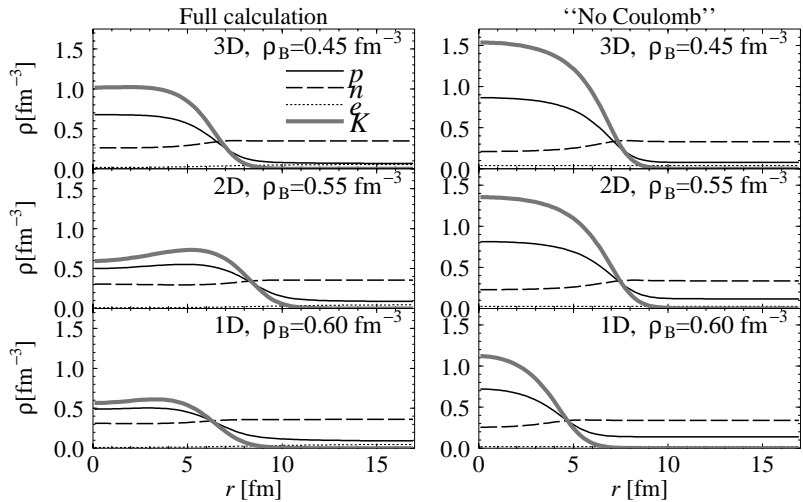}
\end{minipage}
\hspace{\fill}
\begin{minipage}[c]{0.30\textwidth}
  \caption{
  Comparison of density profiles of kaonic matter
  calculated with and without the Coulomb interaction.
  The cell size is not optimized in this comparison.
}\label{figKprofcompare}
\end{minipage}
\end{figure}

We can get more insight about the role of the Coulomb interaction in Fig.~\ref{figKmumu};
left panel in Fig.~\ref{figKmumu} exhibits the full calculation, while 
in the right panel we show the case, when  the electric potential is
discarded in determining the density profile (``No Coulomb'')
and the Coulomb energy,
using the density profile thus determined, is then added to the total energy.
We use ``No Coulomb'' in the same meaning as in the nuclear pastas.
We see that in ``No Coulomb'' case
the pieces of solid curves  lie between two curves given by the bulk
calculation (indicated by ``Gibbs''), where the Gibbs conditions are imposed disregarding the
surface tension and the Coulomb interaction,  
and by the Maxwell
construction (indicated by ``Maxwell'').
The ``Full calculation'' case
is more close to the one given by the Maxwell construction.
As follows from the density profiles shown in Fig.~\ref{figKprofcompare}, the local charge neutrality
is more pronounced in the case  of the ``Full calculation''
(smaller difference of kaon and proton densities).
These results suggest that the Maxwell construction is effectively
meaningful owing to the charge screening effects.

\section{Summary and concluding remarks}\label{Sum}

We have discussed two kinds of the non-uniform structures in nuclear
matter, nuclear pastas at subnuclear densities and the kaon mixed phase
above the nuclear density,
which may arise as consequences of FOPT with many particle species,
and elucidated the charge screening effect.
Using a self-consistent framework based on density functional theory and
relativistic mean fields,
we took into account the Coulomb interaction in a proper way and
numerically solved coupled equations of motion to extract the
density profiles of nucleons, electrons and kaons.

First of all, we have checked how realistic our framework is by studying the bulk
properties of finite nuclei, as well as
the saturation properties of nuclear matter, and found that it can describe
both features satisfactorily. 
One could still improve the consideration by including the gradient
terms of the nucleon densities, which may give a better description
of the nuclear surface.

In isospin-asymmetric nuclear matter with fixed proton to atomic number ratios, 
we have observed the ``nuclear pastas''
with various geometrical structures at sub-nuclear densities. These cases are
relevant for the discussion of the supernova explosions and for the
description of the newly born neutron stars.
The appearance of the pasta structures significantly lowers the energy,
i.e. softens the equation of state, while the energy differences between 
various geometrical structures are rather small. 

By comparing different treatments of the Coulomb interaction,
we have seen that the self-consistent inclusion of the Coulomb
interaction changes the phase diagram.
In particular the region of pasta structure is broader
for ``full'' calculation compared to that with simplified
treatments of the Coulomb interaction which have been used
in the previous studies. 
The effect of the rearrangement of the proton distributions on 
the structures is much more pronounced compared to the effect
of the electron charge screening.
The influence of the charge screening on the equation of state, 
on the other hand, was found to be small.

We have also studied the structure of the nucleon matter in the beta equilibrium.
We have found that  only one type of structures is realized: proton-enriched
droplets embedded in the neutron sea.
No other geometrical structures like rod, slab, etc. appeared.

We have discussed how the geometrical structures manifest in the context
of kaon condensation. Our framework can be extended to include kaons straightforwardly.
%
We have discussed the effect of the charge screening in this case.
Since the kaon mixed-phase appear at high-densities, 
we see that changes are more remarkable than for the ``nuclear pastas''
at subnuclear densities \cite{maru1}.
The density range of the structured mixed
phase is largely limited by the charge screening 
and thereby the phase diagram becomes similar
to that given by the Maxwell construction. 
Although the importance of such a treatment has been demonstrated for the
hadron-quark matter transition \cite{voskre,emaru1}, one of our new 
findings here is that we can
figure out the role of the charge screening effect without introducing
an ``artificial'' input of the surface tension.
In our study we have used the one-boson-exchange interaction for the
$KN$ interaction. 
On the other hand the bulk calculation, where no Coulomb 
interaction nor surface tension is included, 
can not give the mixed phase in the chiral model \cite{PREPL00,MYTT}. 
It should be interesting to study the effects of the charge screening 
to see whether the chiral model is not thermodynamically well-defined.

In application to the newly formed neutron stars like in  supernova explosions, finite temperature and
neutrino trapping effects become important, as well as the dynamics of the
first order phase transition with formation of the structures. 
It would be interesting to extend our
framework to include these effects. 

\section*{Acknowledgments}

This work is partially supported by the Grant-in-Aid for the 21st Century COE
``Center for the Diversity and Universality in Physics'' from 
the Ministry of Education, Culture, Sports, Science and
Technology of Japan. It is also partially supported by the Japanese 
Grant-in-Aid for Scientific
Research Fund of the Ministry of Education, Culture, Sports, Science and
Technology (13640282, 16540246). The work of D.N.V. was also supported in part by the Deutsche
Forschungsgemeinschaft (DFG project 436 RUS 113/558/0-2).

\end{document}